\documentclass{pramana}

\usepackage{graphicx,amsmath,bm}

\begin{document}

\title{The Gravitational Lensing Due to Schwarzchild Black Holes}

\author{Amritansh Mehrotra\textsuperscript{1,2}, R. Kanishka\textsuperscript{1}\textsuperscript{*}}
\affilOne{\textsuperscript{1}Department of Physics, University Institute of Sciences, Chandigarh University, Mohali, Punjab, India 140413\\}
\affilTwo{\textsuperscript{2}Pacif Institute of Cosmology and Selfology (PICS), Sagara, Sambalpur 768224, Odisha, India}


\twocolumn[{

\maketitle

\corres{kanishka.rawat.phy@gmail.com}


\begin{abstract}
Gravitational lensing is a powerful concept in the Astrophysics to study black holes. The gravitational field of a massive object like a galaxy or black hole bends and magnifies the light from a distant object behind it. The Schwarzchild black hole that are the simplest type of black hole, having no charge or angular momentum have been useful to observe the gravitational lensing. In the presented work, Schwarzchild black hole has been simulated keeping the spiral, elliptical, lenticular and irregular galaxies at the background to obtain the gravitational lensing.
\end{abstract}



}]

\doinum{:}
\artcitid{\#\#\#\#}
\volnum{}
\year{2024}
\pgrange{1--8}
\setcounter{page}{1}
\lp{8}

\section{Introduction}
\label{intro}
The Schwarzschild metric \cite{schw} describes the simplest type of black hole-a non-rotating, uncharged black hole \cite{blackhole}, known as a Schwarzschild black hole. It was derived by Karl Schwarzschild in 1916, shortly after Einstein proposed his theory of General Relativity.

The Schwarzschild metric \cite{schw} is a solution to Einstein's field equations in a vacuum, meaning there is no matter present except for a central singularity. The key characteristic of this metric is that it describes spherically symmetric gravitational fields. The Schwarzschild metric remains a foundational solution for understanding black hole physics in general relativity.

A Schwarzschild black hole can form when a massive star collapses under its own gravity. If the core's mass is greater than a limit which is known as Tolman-Oppenheimer-Volkoff limit, no force can counterbalance the gravitational collapse, leading to the formation of a black hole.

The gravitational lensing \cite{citation2}--\cite{citation11} occurs when the gravity of a bigger objects such as galaxy or a cluster of galaxies, bends the light from a more distant object, like another galaxy or a quasar, as it travels toward an observer. This phenomenon can distort, magnify, or even create multiple images of the background object. There are several key uses and applications of gravitational lensing in astrophysics and cosmology. Despite several past and ongoing researches done on gravitational lensing, it has not been able to answer many questions in Astrophysics yet, one of them is the nature of dark matter \cite{citation12}.

The gravitational lensing is a powerful tool for both observational astronomy and theoretical Astrophysics, offering insight into the structure, composition, and evolution of the universe \cite{citation13}--\cite{citation20}. Gravitational lensing have been used to probe dark matter and dark energy, measuring the mass of cosmic structures, probing the early Universe, detecting exoplanets, Hubble constant measurement and, discovering hidden objects \cite{citation21}--\cite{citation26}. Hence gravitational lensing is crucial in Astrophysics. The motivation to do the current work is to simulate the black holes with different galaxies that were kept in the background to find the gravitational lensing. The paper has been organized as follows: the methodology have been described in section~\ref{method} The results have been described in section~\ref{rs} The summary and conclusions have been discussed in the section~\ref{summ}

\section{Methodology}
\label{method}

The methodology lies in solving the Schwarzchild metric for photon trajectories. The Schwarzschild metric in spherical coordinates (t, r, $\theta$, $\phi$) is given by eq~\eqref{eqone} \cite{schw}.

\begin{eqnarray}
ds^{2} = c^{2} \left(1 - \frac{R_{s}}{r} \right)dt^{2} - \left(1 - \frac{R_{s}}{r} \right)^{-1} dr^{2} -r^{2}d\theta^{2} - r^{2}\sin^{2}\theta d\phi^{2} 
  \label{eqone}
  \end{eqnarray}

\noindent Where, $R_{s}$=$\frac{2MG}{c^{2}}$\\
$R_{s}$ is black hole radius\\
M is mass of black hole\\
G = 6.6743 $\times$ $10^{-11}$ $m^{3}$ $kg^{-1}$ $s^{-2}$, is Gravitational constant.\\
c is light's speed in the vacuum.\\

The photon trajectories in the Schwarzschild metric underpin the phenomenon of gravitational lensing, where the light from a far-off source is bent around a big object, acting like a lens. This effect can produce multiple images, magnification, and distortion of the background source. The goal is to find the photon trajectories for gravitational lensing. The photon trajectory in the Schwarzschild metric is governed by the geodesic equation, which describes the motion of a particle (or photon, in this case) in a curved spacetime. Since photons travel along null geodesics, the condition $ds^{2}$=0 applies. For the photons coming from a distance posses a trajectory derived from eq~\eqref{eqone} is given by eq~\eqref{eqtwo} \cite{citationx}.

\begin{eqnarray}
\frac{d^{2}u}{d\phi^{2}} - \frac{3R_{s}u^{2}}{2} + u = 0
  \label{eqtwo}
\end{eqnarray}

\noindent Where, $u(\phi)$ = $\frac{1}{r(\phi)}$\\
Where $\phi$ is the final angle and signifies the bending of photon around the black hole.
In order to see the bending of photons around the black hole, there has been two important angles. One is deviated angle through which the photons get deviated near the black hole and another is seen angle, which actually shows the visualization of photon trajectories to the observer \cite{citationy}. The relation between seen angle and deviated angle has been shown in eq~\eqref{eqthree} \cite{citation1}. 

\begin{eqnarray}
seen~~angle (^{o}) = \pi - \alpha , \nonumber \\
deviated~~angle (^{o}) = \phi + asin(\frac{D \sin(\phi)}{R})
  \label{eqthree}
\end{eqnarray}

\begin{figure*}
\centering\includegraphics[height=.3\textheight]{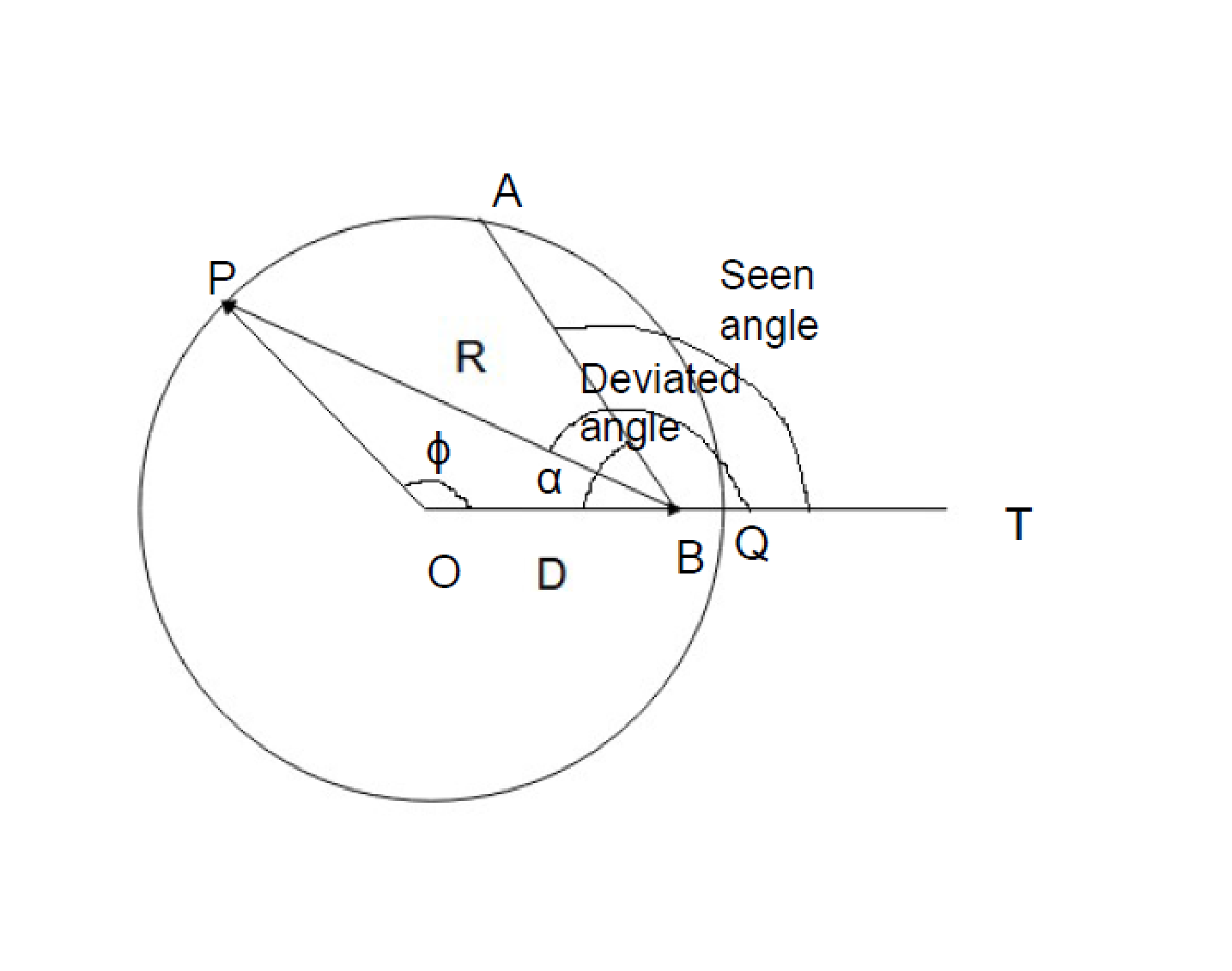}

\caption{The seen and deviated angle by photon trajectories \cite{citationz}.}\label{figOne}
\end{figure*}

\begin{figure*}
\centering\includegraphics[height=.3\textheight]{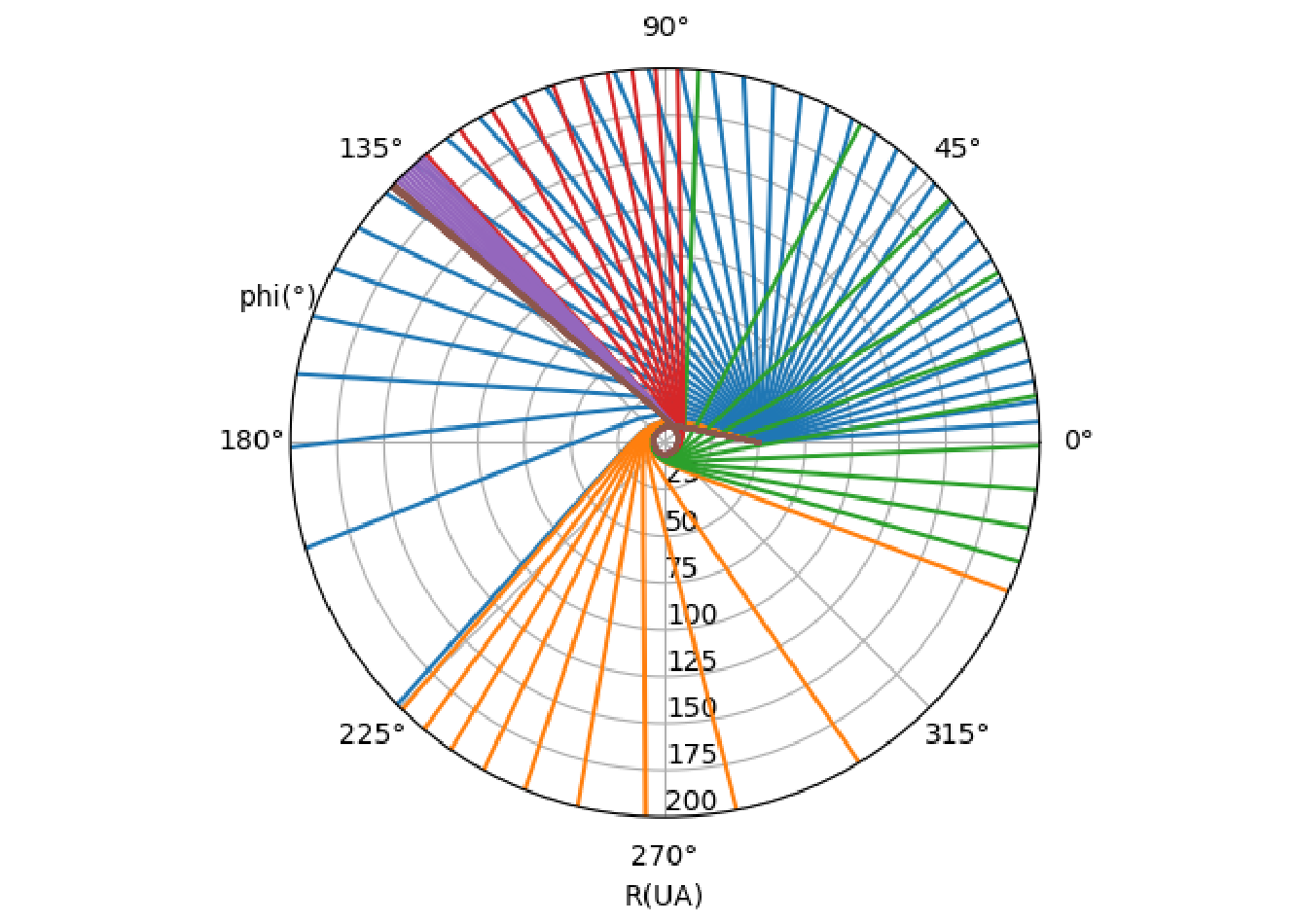}

\caption{The trajectories of photons around black hole.}\label{figTwo}
\end{figure*}

\noindent Where, seen angle and deviated angles are calculated in degrees. \\
$\alpha$ is an initial angle. \\
D = 50 km, is the distance from the black hole .\\
R is the vision radius.\\
These angles and distances have been shown in the figure~\ref{figOne}. The figure~\ref{figTwo} shows the trajectories of photons around the black hole of $R_{s}$ = 12 km for all possible values of $\alpha$. In the next section the results have been discussed.

  \section{Results}
\label{rs}  
The simulation was done using the black hole simulation package \cite{citation1} with the different galaxies that have been discussed in the next sub-section.

\subsection{Simulation of Gravitational Lensing of Black Holes}
\label{el}
The gravitational lensing was observed with four different galaxies. These galaxies are spiral, elliptical, lenticular and irregular. These four types of galaxies are the main galaxies that exist in the Universe. The results obtained for gravitational lensing have been discussed below.

\begin{figure*}
  \centering\includegraphics[height=.2\textheight]{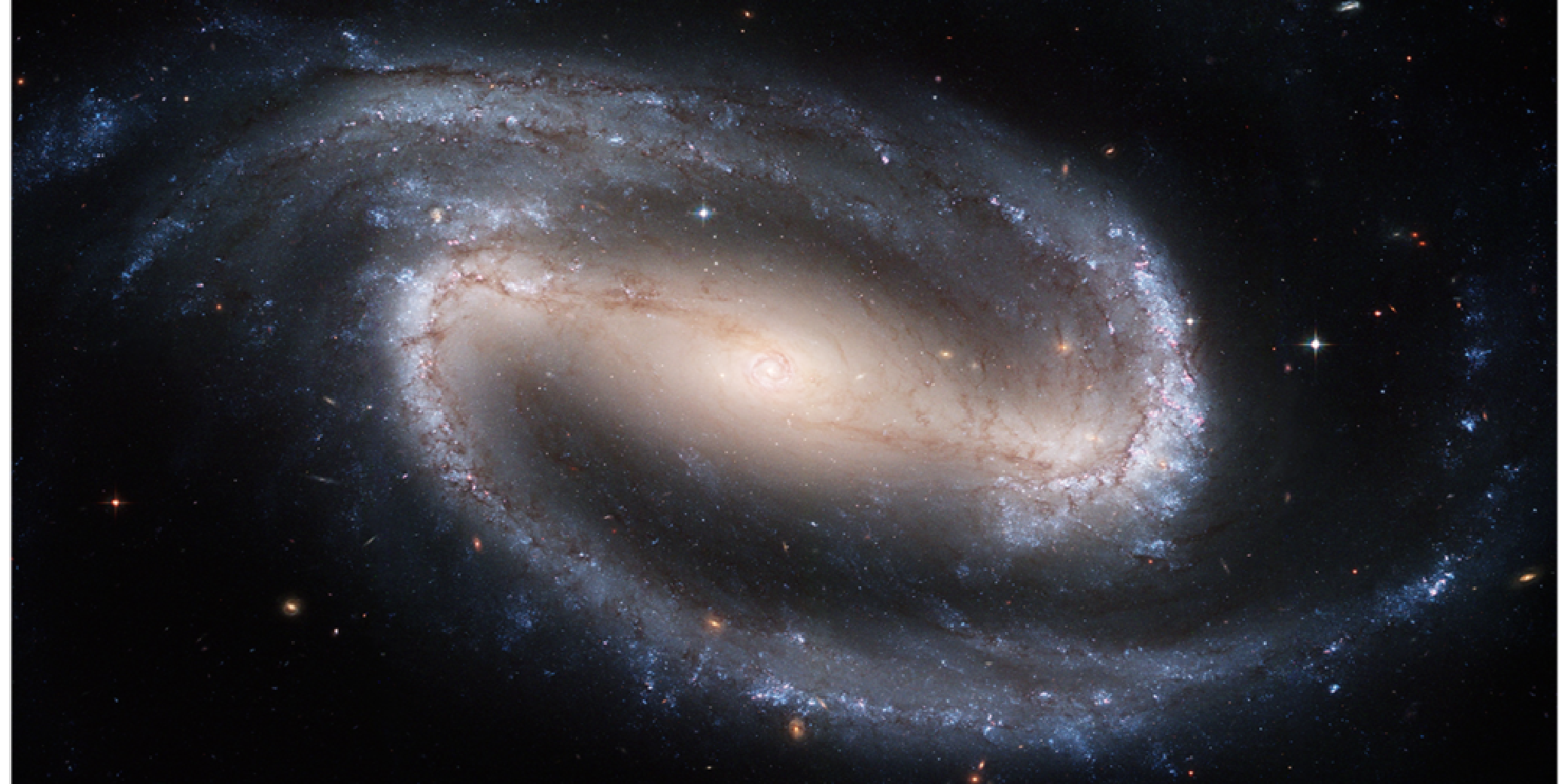}
\centering\includegraphics[height=.2\textheight]{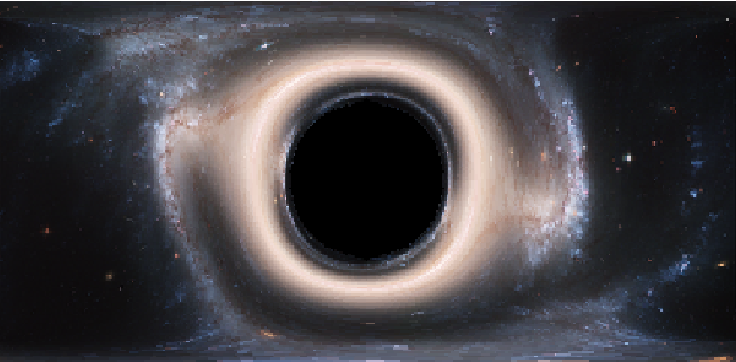}

\caption{The top panel shows spiral galaxy NGC 1300 without the gravitational lensing. The bottom panel shows the gravitational lensing by Schwarzchild black hole with the spiral galaxy NGC 1300 in the background. Note: The black hole of $R_{s}$ = 12 km has been taken here for gravitational lensing.}\label{figThree}
\end{figure*}

\begin{figure*}
  \centering\includegraphics[height=.2\textheight]{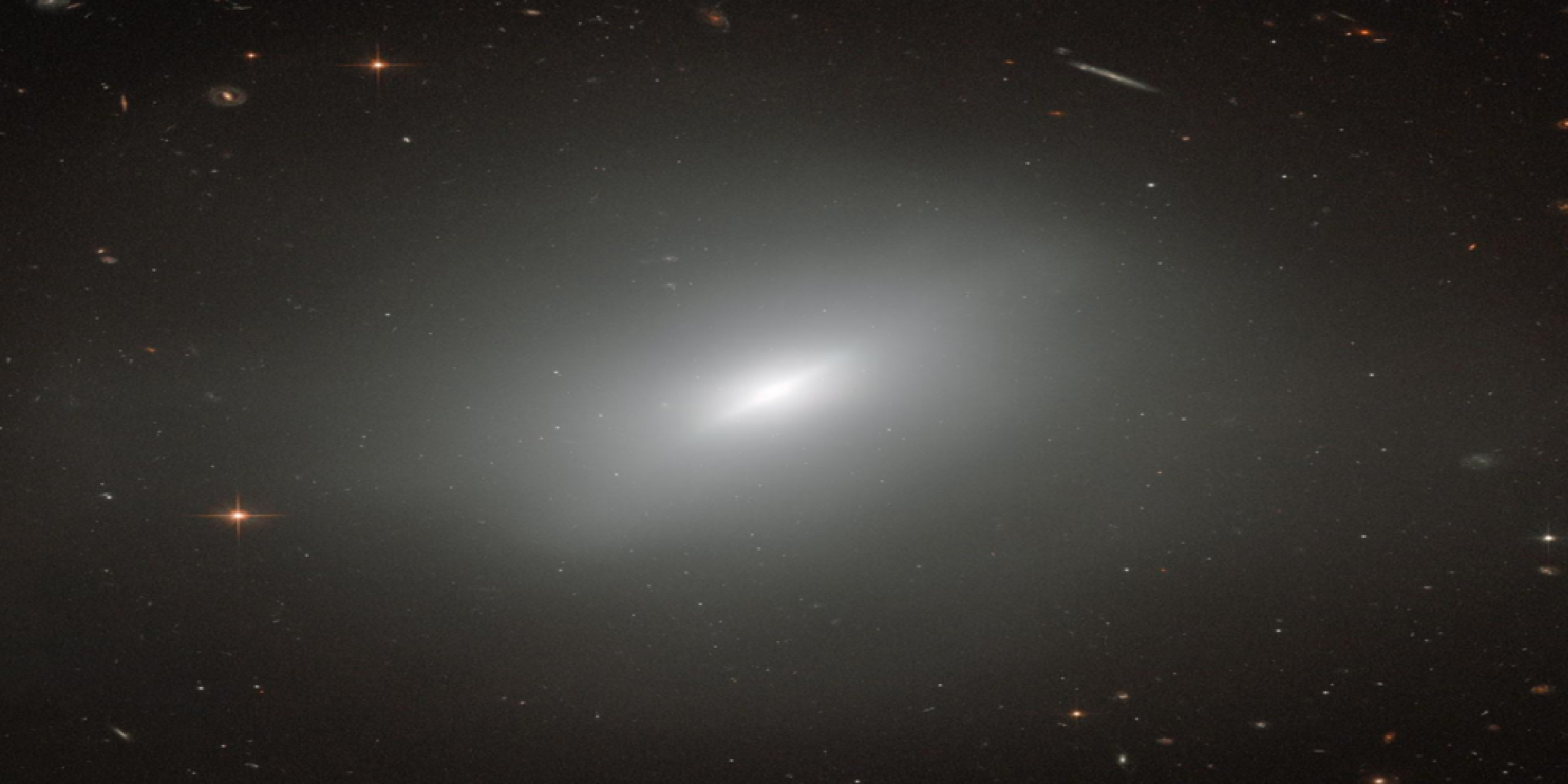}
\centering\includegraphics[height=.2\textheight]{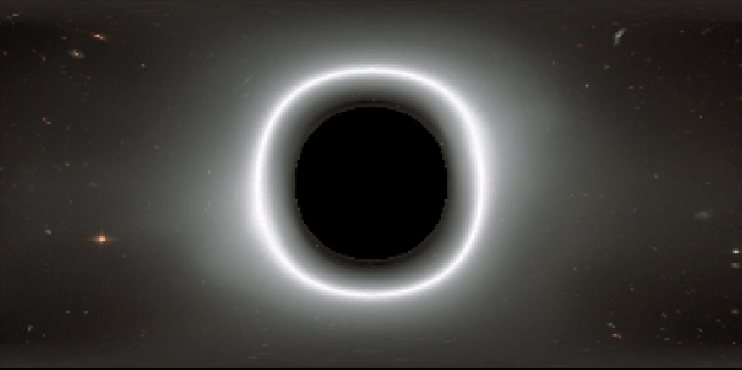}

\caption{The top panel shows elliptical galaxy NGC 3610 without the gravitational lensing. The bottom panel shows the gravitational lensing by Schwarzchild black hole with the elliptical galaxy NGC 3610 in the background. Note: The black hole of $R_{s}$ = 12 km has been taken here for gravitational lensing.}\label{figFour}
\end{figure*}

\begin{figure*}
  \centering\includegraphics[height=.2\textheight]{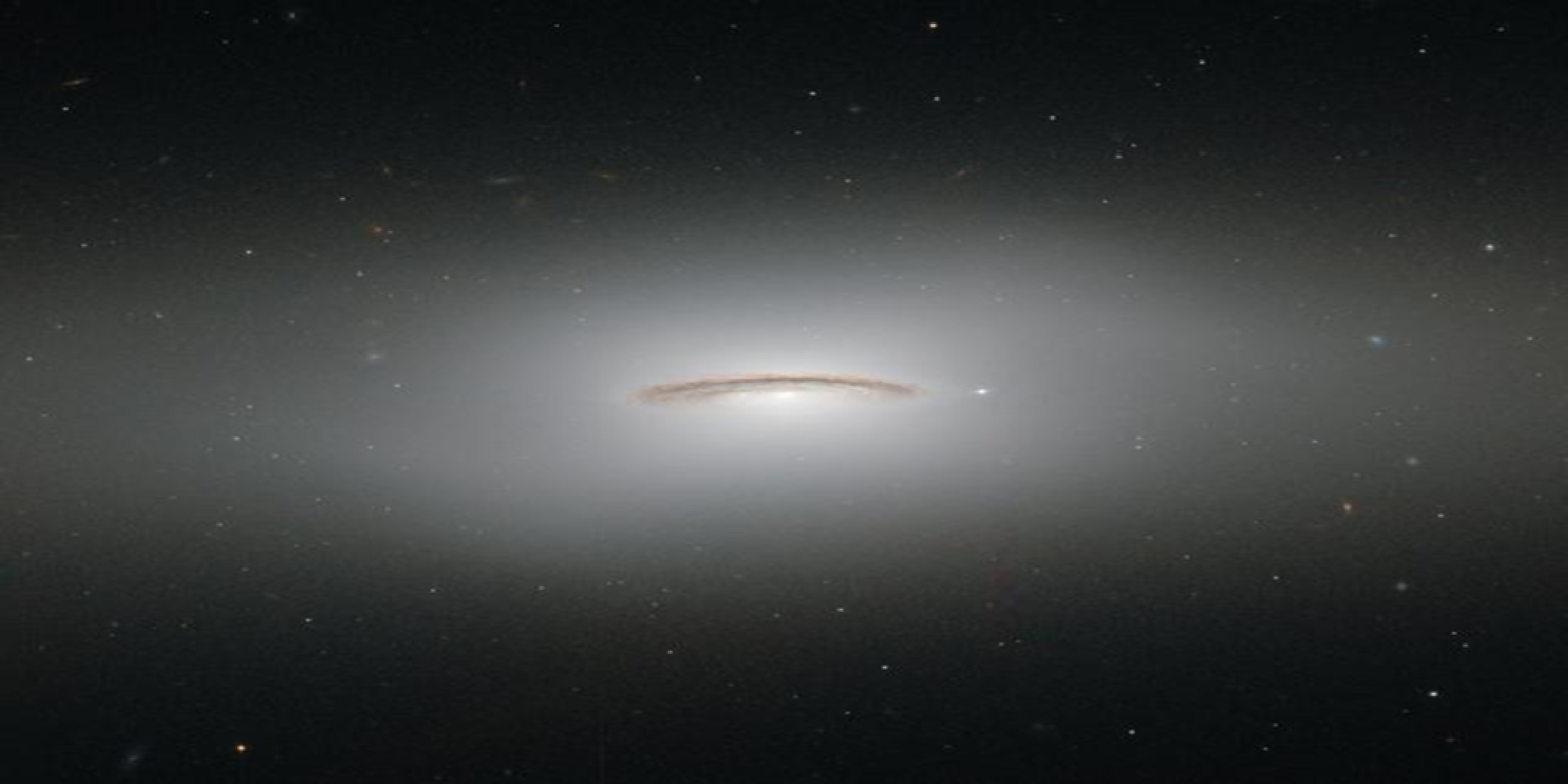}
\centering\includegraphics[height=.2\textheight]{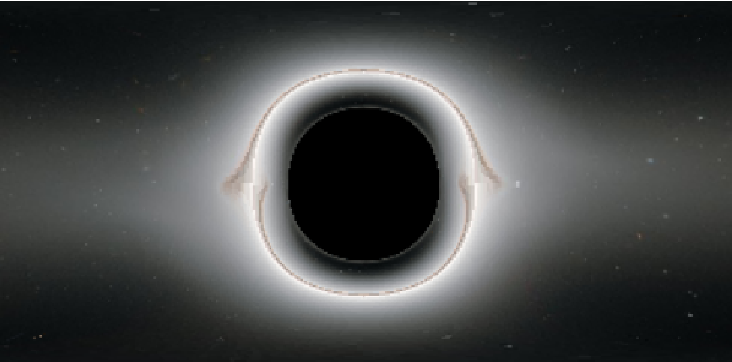}

\caption{The top panel shows lenticular galaxy NGC 4526 without the gravitational lensing. The bottom panel shows the gravitational lensing by Schwarzchild black hole with the lenticular galaxy NGC 4526 in the background. Note: The black hole of $R_{s}$ = 12 km has been taken here for gravitational lensing.}\label{figFive}
\end{figure*}

\begin{figure*}
  \centering\includegraphics[height=.2\textheight]{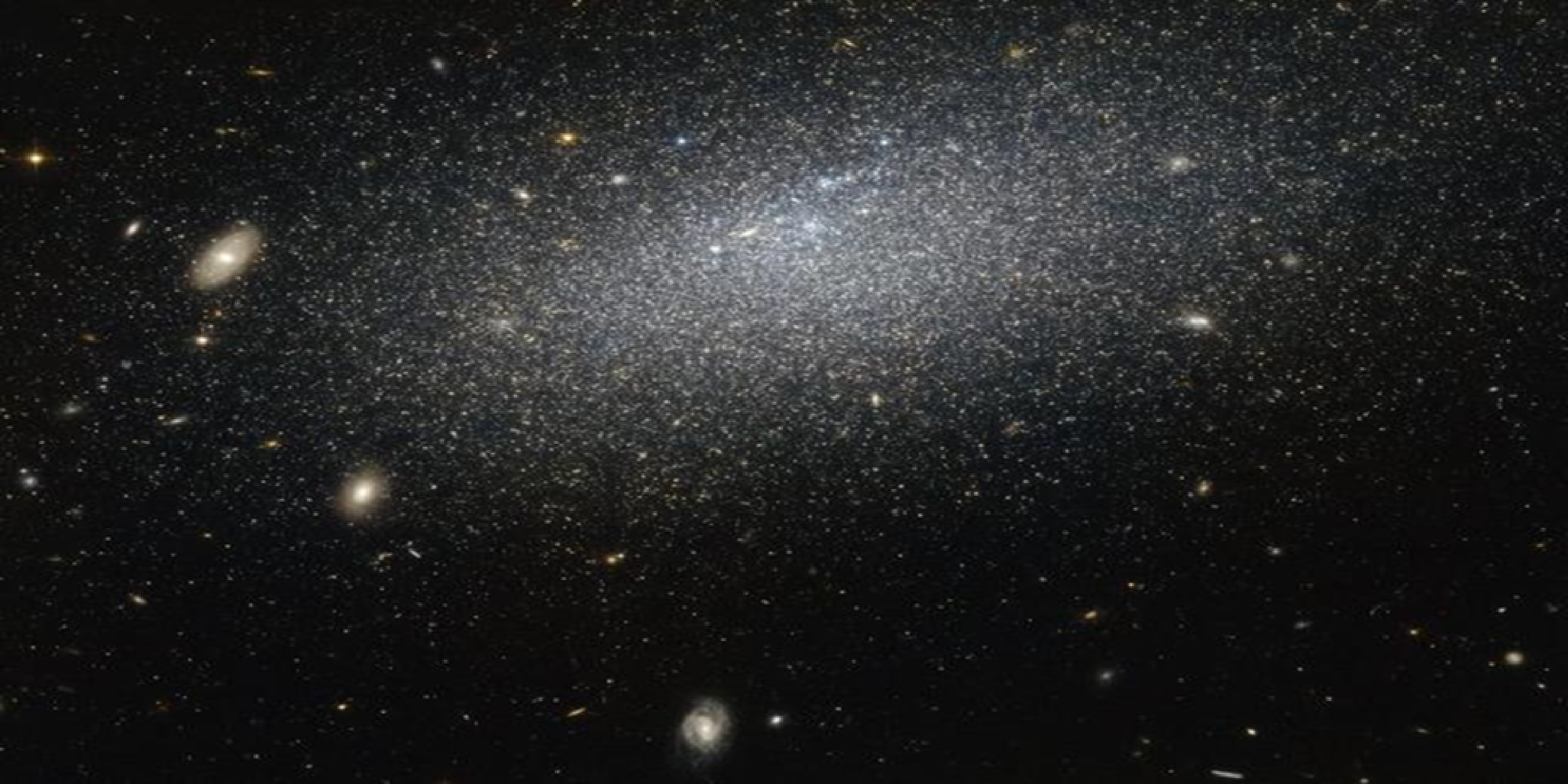}
\centering\includegraphics[height=.2\textheight]{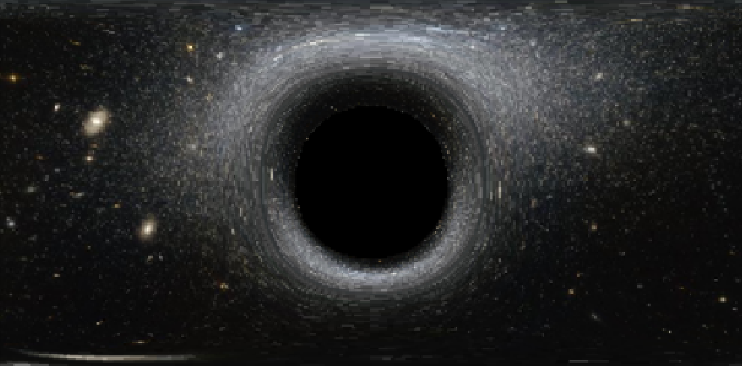}

\caption{The top panel shows irregular galaxy UGC 4879 without the gravitational lensing. The bottom panel shows the gravitational lensing by Schwarzchild black hole with the irregular galaxy UGC 4879 in the background. Note: The black hole of $R_{s}$ = 12 km has been taken here for gravitational lensing.}\label{figSix}
\end{figure*}

\begin{figure*}
\centering\includegraphics[height=.25\textheight]{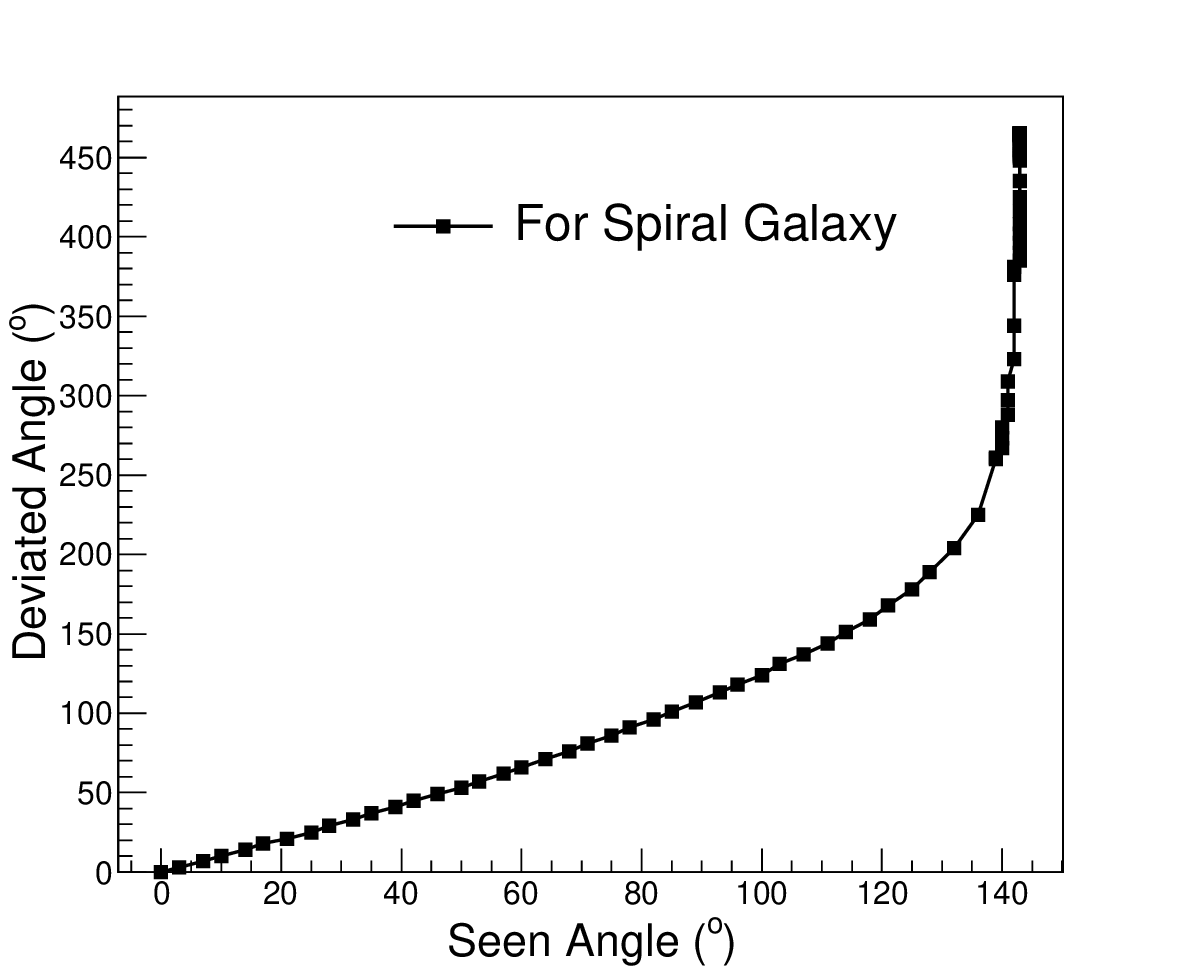}
\centering\includegraphics[height=.25\textheight]{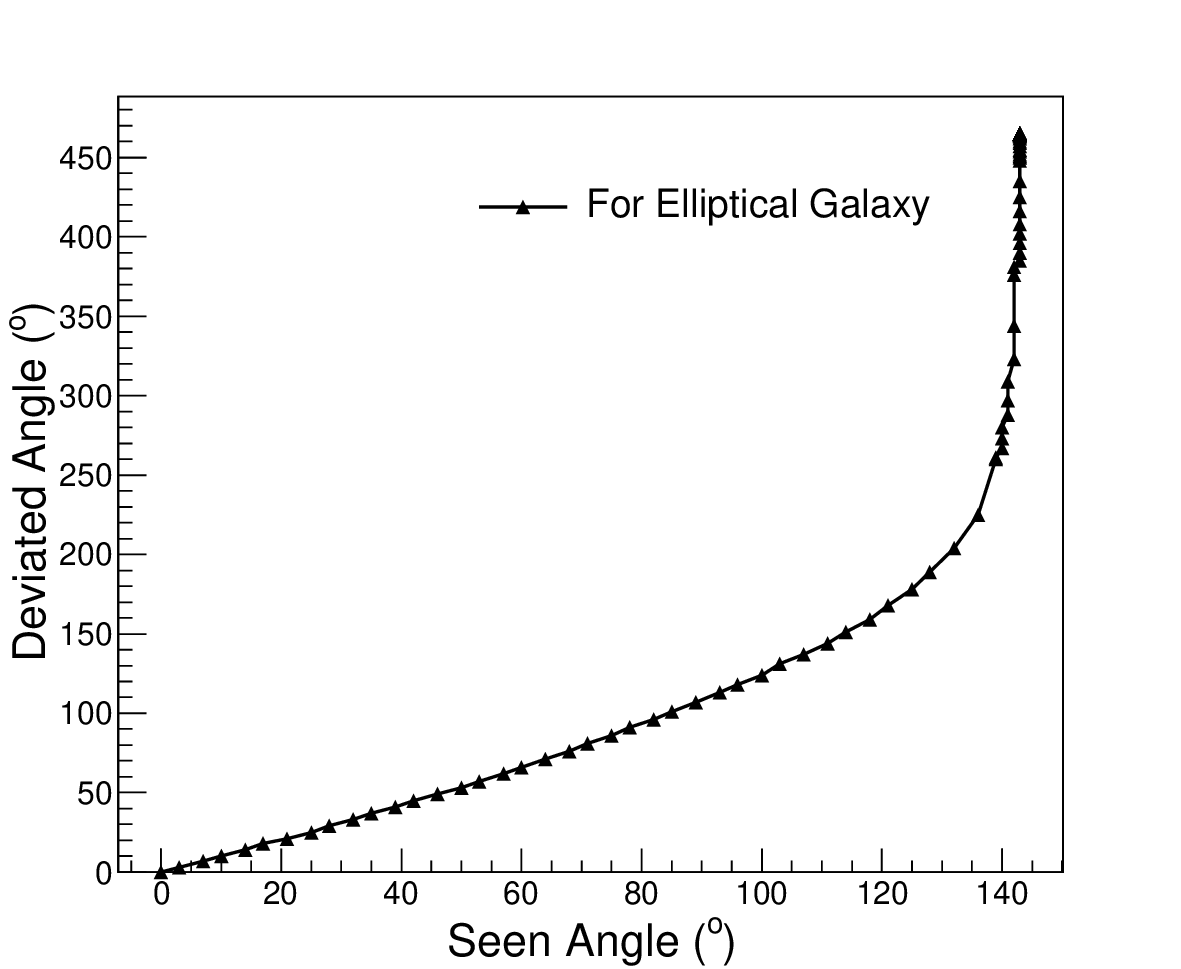}\\
\centering\includegraphics[height=.25\textheight]{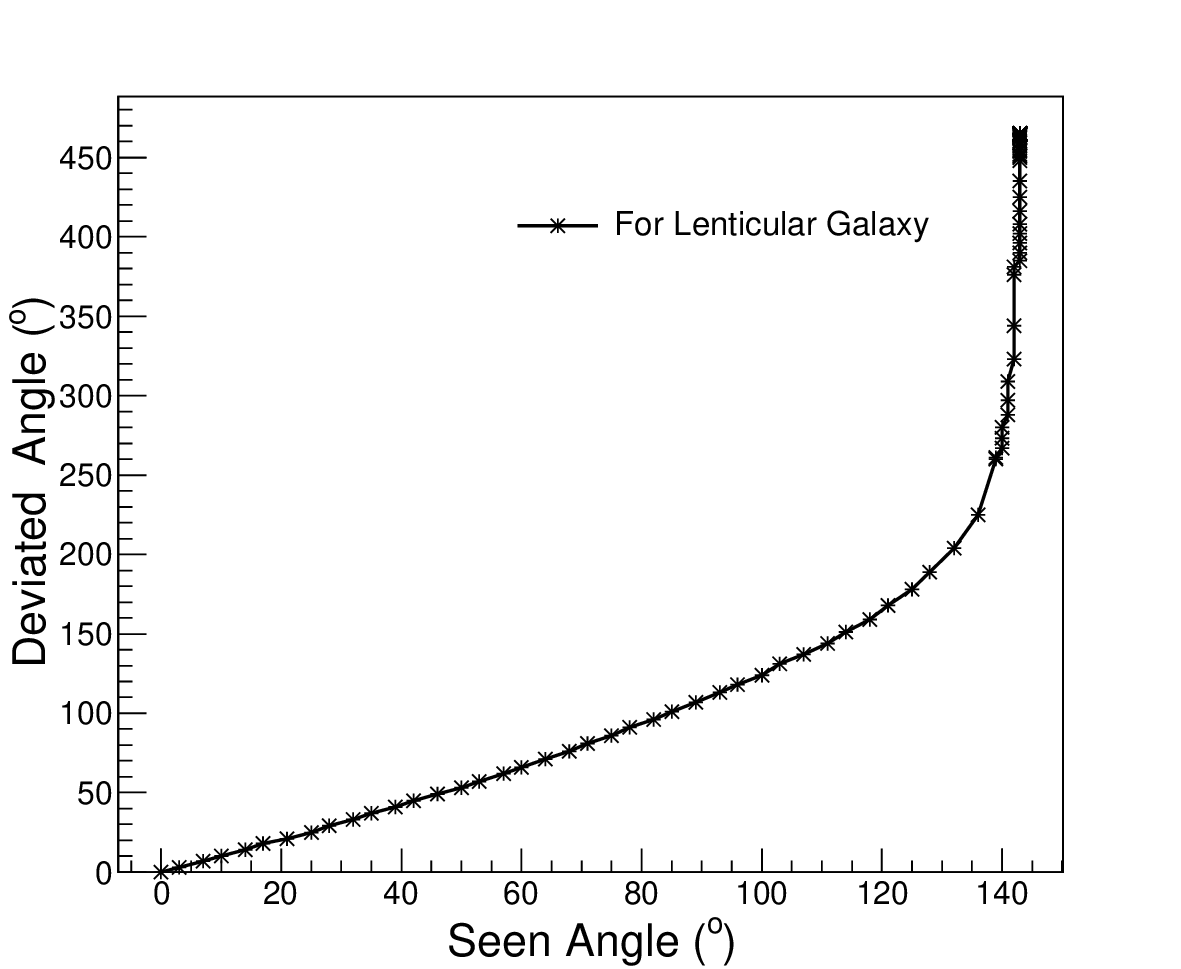}
\centering\includegraphics[height=.25\textheight]{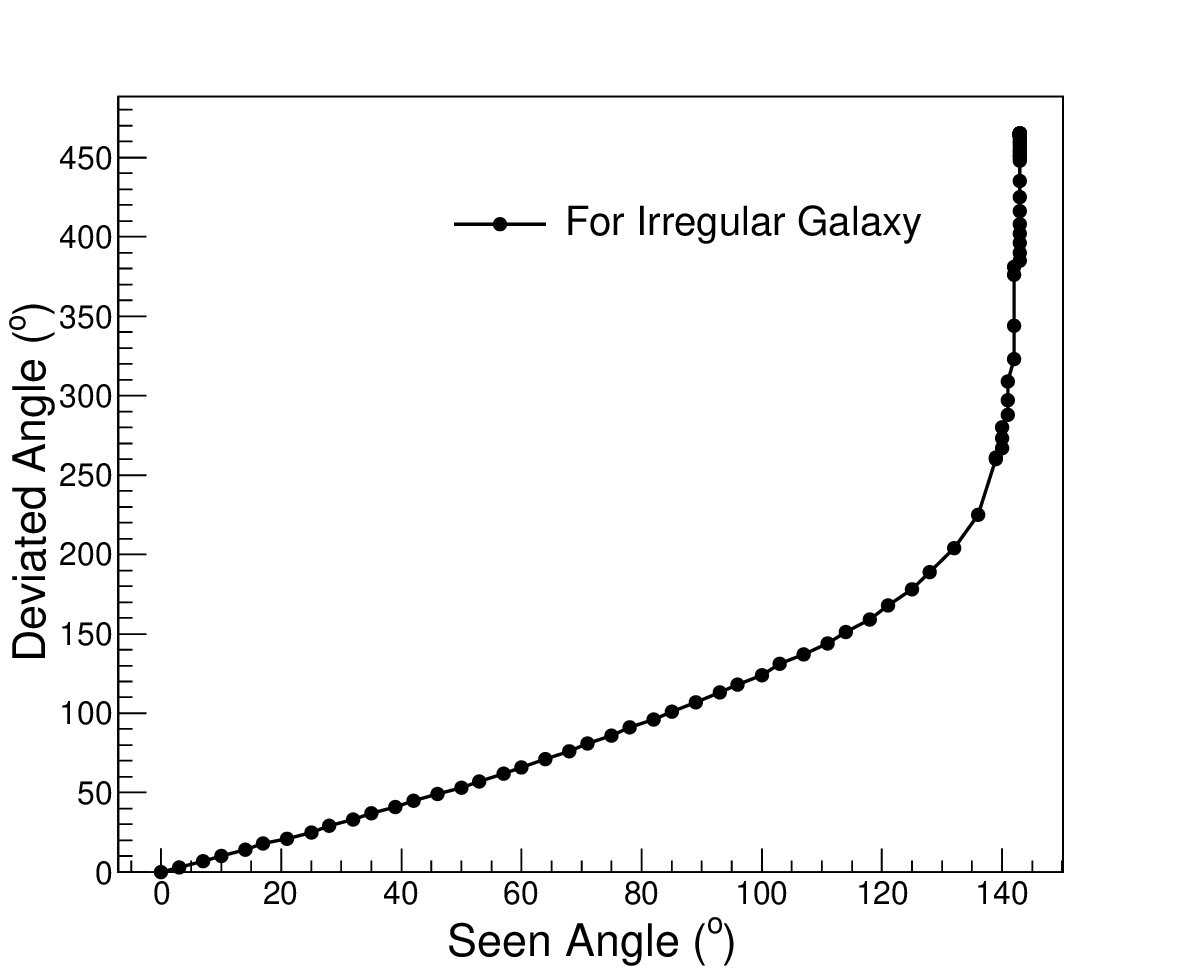}

\caption{The deviated and seen angle with the spiral NGC 1300, elliptical NGC 3610, lenticular NGC 4526 and irregular UGC 4879 galaxies respectively for Schwarzchild black hole of $R_{s}$ = 12 km.}\label{figSeven}
\end{figure*}

The NGC 1300 \cite{spiral} which is a barred spiral galaxy has shown the gravitational lensing in figure~\ref{figThree}. In figure~\ref{figFour}, the elliptical galaxy NGC 3610 \cite{elliptical} used to be present in constellation Ursa major was considered for gravitational lensing. Figure~\ref{figFive} shows the gravitational lensing with NGC 4526 \cite{lenticular} which is a lenticular galaxy. In figure~\ref{figSix}, UGC 4879 \cite{irregular} which is an irregular dwarf galaxy has shown the gravitational lensing. The top panel of figure~\ref{figThree}--figure~\ref{figSix} shows spiral galaxy NGC 1300, elliptical galaxy NGC 3610, lenticular galaxy NGC 4526, and irregular galaxy UGC 4879 respectively without the gravitational lensing. The bottom panel of figure~\ref{figThree}--figure~\ref{figSix} demonstrates spiral galaxy NGC 1300, elliptical galaxy NGC 3610, lenticular galaxy NGC 4526, and irregular galaxy UGC 4879 respectively with the gravitational lensing by Schwarzchild black hole. The black hole of $R_{s}$ = 12 km of mass = 4.067 $\times$ the mass of sun has shown gravitational lensing. It was observed that only spiral NGC 1300, elliptical galaxy NGC 3610 and lenticular galaxy NGC 4526 have shown the ring around the black hole where irregular galaxy UGC 4879 has not. This is due to the fact that spiral, elliptical and lenticular galaxies have particular shapes whereas irregular galaxies have not. Figure~\ref{figSeven} demonstrates the deviated and seen angle with the spiral NGC 1300, elliptical NGC 3610, lenticular NGC 4526 and irregular UGC 4879 galaxies respectively for Schwarzchild black hole of $R_{s}$ = 12 km  as discussed in section~\ref{method} It was observed that the seen angle increases exponentially with the deviated angle. This shows the bending of photons around the Schwarzchild black hole as discussed in section~\ref{method} It was also observed that the deviated and seen angle by the different galaxies show the same behaviour. The next section summarizes and concludes the discussion.

\section{Summary and Conclusions}
\label{summ}
A Schwarzschild black hole represents the simplest type of black hole, described by a solution to Einstein's field equations in general relativity. It is characterized by having mass but no electric charge or angular momentum. The Schwarzschild black hole lays a foundation for studying black holes in general relativity. One of the important property of black hole is the gravitational lensing. The gravitational lensing is a powerful natural telescope that allows us to study distant and faint objects. The light from a distant object is bent as it passes near a bigger object, such as a black hole or a galaxy. In the current work the Schwarzschild black hole was chosen to observe gravitational lensing with four different types of galaxies. The photons coming from these galaxies bend around the black hole and show lensing effect. It was found that only spiral NGC 1300, elliptical galaxy NGC 3610 and lenticular galaxy NGC 4526 have shown the ring around the black hole, that is because these galaxies have particular shapes.

\section*{Acknowledgement}
The authors thank the department of physics, Chandigarh University for help and support.

\end{document}